\documentclass[12pt,preprint]{aastex}
\newcommand{\etal}{\mbox{\it et al.~}}
\newcommand{\ms}{\mbox{m s$^{-1}~$}}
\newcommand{\ks}{\mbox{km s$^{-1}~$}}

\newcommand{\kse}{\mbox{km s$^{-1}$}}
\newcommand{\mse}{\mbox{m s$^{-1}$}}
\newcommand{\msun}{M$_{\odot}~$}
\newcommand{\mstar}{M$_{\star}~$}
\newcommand{\msune}{M$_{\odot}~$}
\newcommand{\rsun}{R$_{\odot}~$}
\newcommand{\rstar}{R$_{\star}~$}

\newcommand{\mjup}{M$_{\rm JUP}~$}

\newcommand{\mjupe}{M$_{\rm JUP}$}

\newcommand{\msini}{$M \sin i~$}
\newcommand{\vsini}{$v \sin i~$}

\newcommand{\chisq}{$\sqrt{\chi_{\nu}^2}~$}

\newcommand{\teff}{$T_{\rm eff}~$}

\newcommand{\logg}{${\rm \log g}~$}

\received{}
\accepted{}

\slugcomment{submitted ApJ}

\shortauthors{Fischer \& Laughlin}
\shorttitle{N2K}
\begin{document}

\title{A Hot Saturn Planet Orbiting HD 88133, from the N2K Consortium\altaffilmark{1}}
\author{Debra A. Fischer\altaffilmark{2}, 
Greg Laughlin\altaffilmark{3},
Paul Butler\altaffilmark{4},
Geoff Marcy\altaffilmark{5},
John Johnson\altaffilmark{5},
Greg Henry\altaffilmark{6},
Jeff Valenti\altaffilmark{7},
Steve Vogt\altaffilmark{3},
Mark Ammons\altaffilmark{3},
Sarah Robinson\altaffilmark{3},
Greg Spear\altaffilmark{3},
Jay Strader\altaffilmark{3},
Peter Driscoll\altaffilmark{2},
Abby Fuller\altaffilmark{2},
Teresa Johnson\altaffilmark{2},
Elizabeth Manrao\altaffilmark{2},
Chris McCarthy\altaffilmark{2},
Melesio Mu\~noz\altaffilmark{2},
K. L. Tah\altaffilmark{2},
Jason Wright\altaffilmark{5},
Shigeru Ida\altaffilmark{8},
Bun'ei Sato\altaffilmark{9},
Eri Toyota\altaffilmark{9},
Dante Minniti\altaffilmark{10}}

\email{fischer@stars.sfsu.edu}

\altaffiltext{1}{Based on observations obtained at the W. M. Keck Observatory, 
which is operated by the University of California and the California Institute of 
Technology. Keck time has been granted by NOAO and NASA.}

\altaffiltext{2}{Department of Physics \& Astronomy, San Francisco State University, 
San Francisco, CA  94132; fischer@stars.sfsu.edu}

\altaffiltext{3}{UCO/Lick Observatory, University of California at Santa Cruz, 
Santa Cruz, CA 95064}

\altaffiltext{4}{Department of Terrestrial Magnetism, Carnegie Institute of 
Washington DC, 5241 Broad Branch Rd. NW, Washington DC, USA 20015-1305}

\altaffiltext{5}{Department of Astronomy, University of California, Berkeley, CA USA 94720}

\altaffiltext{6}{Center of Excellence in Information Systems, Tennessee State University, 
330 10th Avenue North, Nashville, TN 37203; Also Senior Research Associate, Department of 
Physics and Astronomy, Vanderbilt University, Nashville, TN 37235}

\altaffiltext{7}{Space Telescope Science Institute, 3700 San Martin Dr., Baltimore, MD 21218}

\altaffiltext{8}{Tokyo Institute of Technology, Ookayama, Meguro-ku, Tokyo 152-8551, 
Japan; and University of California Observatories, Lick Observatory, University of 
California, Santa Cruz, CA 95064}

\altaffiltext{9}{Graduate School of Science and Technology, Kobe University, 
1-1 Rokkodai, Nada, Kobe 657-8501, Japan}

\altaffiltext{10}{Department of Astronomy, Pontificia Universidad Catolica, Avenida
Vicu\~na Mackenna 4860, Casilla 306 Santiago 200, Chile}

\begin{abstract}
The N2K consortium is carrying out a 
distributed observing campaign with the Keck, Magellan and
Subaru telescopes, as well as the
automatic photometric telescopes of Fairborn Observatory, in order 
to search for short-period gas giant planets around metal-rich stars. 
We have established a reservoir of more than 14,000 main sequence and subgiant 
stars, closer than 110 pc, brighter than V=10.5 and with $0.4 < B-V < 1.2$.
Because the fraction of stars with planets is a 
sensitive function of stellar metallicity, a broadband photometric 
calibration has been developed to identify a subset of 2000 stars
with [Fe/H] $> 0.1$ dex for this survey.  
We outline the strategy and report the detection of a planet  
orbiting the metal-rich G5IV star HD 88133 with a period of 3.41 days, semi-velocity 
amplitude, K=35.7 \mse and \msini = 0.29 \mjupe. 
Photometric observations reveal that HD~88133 is constant on the 
3.415-day radial velocity period to a limit of 0.0005 mag.  Despite 
a transit probability of 19.5\%, our photometry rules out the shallow 
transits predicted by the large stellar radius.

\end{abstract}


\section{Introduction}

The Doppler Radial Velocity (RV) technique is an 
effective tool that has resulted in the detection of 
136 extrasolar planets over the last decade.
Ongoing Doppler projects in the U.S. and Europe are currently surveying
almost 3000 of the closest and brightest stars ($V<8$). These surveys now 
detect Jupiter-like extrasolar planets in wider orbits analogous to our own 
solar system (Marcy \etal 2002) and Neptune-mass planets with orbital 
periods of a few days (Butler \etal 2004, McArthur \etal 2004, Santos \etal 2004).
Because planets with short orbital periods are so easily detected 
with Doppler RV measurements, virtually all planets with \msini $> 0.5$ \mjup 
and orbital periods less than 14 days (the so-called ``hot Jupiters'') have 
already been harvested from current Doppler surveys. 

The observation of a hot Jupiter planet transiting the star HD 209458 (Charbonneau \etal 2000, 
Henry \etal 2000) provided an important direct detection of an extrasolar planet. 
The quality of physical information for HD 209458b exceeds that of any other 
known exoplanet because the intrinsic brightness of the host star enables the efficient
collection of high-precision (3 $m s^{-1}$) radial velocities which allow
precise determination of the mass and orbital parameters (e.g. Butler et
al 1996), as well as high temporal resolution HST photometry (Brown \etal
2001) and spectroscopic observations of 
the atmospheric constituents in the planet atmosphere (Charbonneau \etal 2002, 
Vidal-Madjar \etal 2003).  The intrinsic brightness of 
HD 209458 also contributed to the precision with which stellar characteristics 
(\teff, metallicity, \vsini, \mstar, \rstar) could be derived and an accurate parallax 
measurement is available from Hipparcos.   Since only
the ratio of the planet radius to the stellar radius can be determined, knowledge of the 
stellar parameters is key to deriving the radius and density of the transiting 
planet. 

The recent spate of discoveries of planets transiting faint stars
(Alonso \etal 2004, Bouchy \etal 2004, Konacki \etal 2003, Pont \etal
2004) indicates that HD 209458b is in fact a rather anomalous planet, with
a radius 40\% larger than expected both from theory. The radius of HD 209458 
is also apparently larger than the radii of the other transiting planets. The radius discrepancy
observed for HD 209458b has spawned controversy regarding the nature of
the mechanism responsible for the planet's bloated condition (Guillot \&
Showman, 2002, Baraffe \etal 2003, Burrows \etal 2003, Bodenheimer, Laughlin \& Lin 2003).

Since the detection of HD 209458b, more than twenty photometric transit searches 
have been started (Horne 2003).  Unfortunately, these surveys cannot detect the 
90\% of hot Jupiter planets that do not transit their host stars.
Non-transiting hot Jupiters are important objects in their own right and the detection 
of a large number of these objects could provide powerful constraints on theories 
of planet formation and evolution.  For example, 
these planets can exhibit non-Keplerian interactions with sibling planets on 
timescales of just a few years.  In addition, the eccentricity 
distribution of short-period planets could reveal whether hot Jupiters have solid cores. 
The detection of a large number of new hot Jupiters could also shed light on 
the puzzling concentration of planets in the period range between
roughly 2.5 d and 3.5 d (9 out of 20 objects with $P<10\,{\rm d})$. 
This orbital pile-up at 3 days is seemingly 
at odds with the detection of 3 out of 4 transiting planets with orbital periods 
of about 1.5 days (``very hot Jupiters'') from the OGLE survey (Konacki \etal 2004). There is 
no observational bias against the Doppler detection of $P=1.5$~day planets by 
the Doppler technique, suggesting that these very hot Jupiters must be 
much less common than gas giant planets in 3 day orbits.

\section{N2K Consortium}
Current Doppler surveys have identified about 20 extrasolar planets with orbital periods 
shorter than 14 days, however after an initial burst of discoveries, 
the rate of hot Jupiter detections has trailed off.  The known set of hot Jupiters is 
a stagnant set; virtually all of these easily-detected planets 
have been harvested from current RV surveys.  In order to find 
a substantial number of new hot Jupiters, we have established a consortium 
of U.S., Chilean and Japanese astronomers to carry out a distributed observing 
program.  Using the Keck, Magellan and Subaru telescopes, we will observe the 
next 2000 (N2K) closest, brightest, and most metal-rich FGK stars not on current Doppler surveys.

We first established a reservoir of more than 14,000 main sequence and subgiant stars 
drawn from the Hipparcos catalog (ESA 1997).   Considerable information is available 
for this large aggregate of candidate stars including B and V colors, 
2MASS $JHK$ photometry, parallaxes, luminosity, proper motions, photometric variability, 
and information regarding the presence of companions.  Our reservoir stars 
all have $0.4 < B - V < 1.2$ and $V < 10.5$ and all are closer than 110 pc.

The N2K program actively tracks stars as they are drawn out of the reservoir and 
distributed to a particular telescope for an observing run.   At each telescope, 
we use an iodine cell to provide a wavelength reference spectrum and our standard Doppler 
pipeline to analyze all stars identically.
A spectral synthesis modeling pipeline has also been established at each 
telescope so that spectroscopic analysis (Valenti \& Fischer 2004) 
is carried out identically for all stars.  Our strategy is to observe a set of 
star three times over a period of a few days to detect short period RV variations
consistent with a hot Jupiter.  Monte Carlo simulations show that with a Doppler 
precision of 7 \ms, more than 90\% of the planets with \msini $> 0.5$ \mjup and orbital periods 
between 1.2 and 14 days will show $> 3 \sigma$ RV scatter.  Spectroscopic binaries 
typically show RMS scatter of several hundred \ms on this timescale and can be
immediately dropped.  In fact, our Doppler velocities generally have a better-than-target 
precision 
of 4-5 \ms and any stars with significant RV variations obtain immediate RV and 
photometric follow-up at the telescopes available to the consortium members.  
Over the next 2 years, the N2K consortium should 
detect $\sim$60 new planets with \msini $> 0.5$ \mjup and orbital 
periods shorter than 14 days and flag stars having longer period 
exoplanets.   

All stars with significant short-term RV variations will also be observed
with the automatic photometric telescopes (APTs) at Fairborn Observatory
(Henry 1999, Eaton, Henry \& Fekel 2003).  As outlined in Butler et al.
(2004), the precise photometric measurements from the APTs are useful
for establishing whether the RV variations are caused by photospheric
features such as spots and plages or by planetary-reflex motion.  Also,
the efficiency and flexibility of the APTs make them ideal for searching
for possible transits that would allow the determination of planetary
radii and true masses.  The high precision of the APT photometry renders
transits of hot Jupiters easily visible, even for the subgiant stars in
our sample with radii of up to 2.0 $R_{\sun}$ (e.g., Henry 2000; \S5 below).

\subsection{Synthetic Templates} 
In the past 6 months we have been testing the use of 
synthetic templates for our Doppler analysis pipeline. The advantage to a synthetic 
template is that it eliminates one RV observation, traditionally taken without iodine. 
To create the synthetic template we first divide the stellar observation 
(with I2) by a featureless B star observation (also with I2).   This division is never 
perfect; residual 1\% iodine lines are left in the spectrum and our first tests showed
that this resulted in unacceptable drifts in the wavelength scale as the barycentric 
velocities change with time.  To try to eliminate this velocity drift, we morphed  
the National Solar Observatory (NSO) solar spectrum (Wallace \etal 1993) to 
match our iodine-divided stellar spectra.
The morphing process involves rotationally broadening the solar absorption lines 
and then globally rescaling with a pseudo optical depth.  Additional fine-tuning is 
accomplished by multiplying the morphed spectrum by the remaining smoothed residuals. 
The morphed-NSO templates provide reliable short-term
RV precision of about 7 \ms and help to identify spectroscopic binaries that should be dropped
from the N2K program.  However, the standard Doppler technique employing an extra
template observation still yields the highest precision and is used for stars that 
warrant long-term RV follow up.

\section{Targeting Metal-Rich Stars}

Butler \etal (2001) have shown that 0.75\% of the stars on 
current Doppler surveys have hot Jupiter companions with orbital periods between 
3--5 days.   However, we expect a higher detection rate because we are 
exploiting an observed correlation between stellar metallicity and the rate of occurrence
of gas giant planets. Fischer \& Valenti (2004) show that stars with [Fe/H] $>0.2$ have at 
least 3 times as many extrasolar planets as solar metallicity stars and we will bias our 
sample with metal-rich stars.  In addition,  
our observing strategy will flag orbital periods out to 14 days so 
our detectable parameter space extends beyond the orbits considered to be hot Jupiters
by Butler \etal (2001).  We expect to find close-in gas giant 
planets around at least 3\% of the stars surveyed by the N2K consortium. 

Of key importance to the success of this project, a photometric calibration 
using broadband filters was developed (Ammons \etal 2004) to provide 
\teff and metallicity estimates for every star in our reservoir 
using Tycho $BV$ and 2MASS $JHK$ photometry.  Our ability to produce a 
Hipparcos-2MASS-metallicity calibration was made possible by our 
possession of a high-quality ``training set'' of spectroscopic metallicities 
for more than 1100 FGK stars (Valenti \& Fischer 2004).  
Figure 1 shows a histogram of the subset of super-solar metallicity stars from 
our reservoir sample.  This set of more than 2000 stars with [Fe/H] $> 0.1$ dex
comprises the core target sample for the N2K program.   We are also obtaining
low resolution spectroscopy to check the broadband photometry metallicity 
calibration.  The low resolution spectroscopy 
employs a set measured spectral line equivalent widths, recently 
re-calibrated (Robinson \etal 2004) using the same training set of 
spectroscopic metallicities (Valenti \& Fischer 2004).  These indices provide 
a spectroscopic abundance that agrees to better than 0.1 dex (Figure 2).

\section{The First N2K Run}

The first set of N2K stars were observed during a 3 night observing run at 
Keck in the 2004A semester.  After a single observation on the first 
night, we determined the chromospheric activity of the observed 
stars (Wright \etal 2004, Baliunas \etal 1997,
Noyes \etal 1984) and ran the spectra through our 
spectral synthesis pipeline to determine metallicity, \logg,  
\vsini and to identify double-lined spectroscopic binaries (SB2s).
As a result of this ``morning after'' screening, 12 stars were identified 
as SB2's or rapid rotators and were dropped before a second observation 
was obtained.  All information regarding every observed 
star, including RV measurements, information from spectral synthesis modeling,
information about the presence of stellar companions and chromospheric 
activity measurements will be appearing in a subsequent catalog paper. 

We obtained three or more RV measurements for 211 stars at Keck in the 
2004A semester with a typical precision of 4 \ms for stars with standard 
templates and 7 \ms for stars with synthetic templates.  The RV variations 
fell into three categories:  1) 148 stars showed less than $3 \sigma$ RV scatter 
and were retired from the N2K program, 2) 19 stars were spectroscopic binaries 
with RV variations $> 500$ \ms - these spectroscopic binaries were also 
retired from the N2K program, 3) 35 stars (16\% of the 
211 star sample) showed RMS scatter $> 3 \sigma$ consistent with the presence 
of a short-period planet. 

The distribution of RMS scatter for these stars is shown in Figure 3.
The cross-hatched bins in Figure 3 represent the 35 stars having 
radial velocity scatter between 20 -- 80 \mse.  These are the 
stars that warrant RV follow-up to 
search for short-period planets.  One of these stars, HD 88133, showed an 
initial RMS scatter of 24 \mse.  This star now has 
enough RV measurements to characterize an orbiting hot Jupiter planet, 
described below. 

Spectral synthesis modeling was carried out for all 211 stars yielding 
[Fe/H], \teff, \logg and \vsini with uncertainties of 0.04 dex, 23K, 
0.05 dex and 0.3 \ms respectively, as discussed in Valenti \& Fischer (2004).  
Figure 4 shows that the spectroscopic metallicities agree well with the 
broadband metallicity estimates (Ammons \etal 2004), 
confirming that the broadband photometric calibration is an excellent way to identify metal-rich 
target stars for the N2K program. 

\section{HD 88133}

HD 88133 is a G5 IV star with $V=8.0$ and $B - V = 0.81$. The Hipparcos parallax 
(ESA 1997) of $13.43~ mas$ places the star at a distance of 74.5 pc with an absolute
visual magnitude, $M_V = 3.65$.  Our spectroscopic analysis yields \teff = 5494 $\pm 23$ K,
[Fe/H] = 0.34 $\pm 0.04$, \logg = 4.23 $\pm 0.05$ and \vsini = 2.2 \kse.  From the 
bolometric luminosity and \teff we derive a stellar radius of 1.93 \rsun and 
evolutionary tracks of Girardi \etal provide a stellar mass estimate of 1.2 \msune.
The chromospheric activity of HD~88133 is $S_{HK} = 0.138$ and $\log R'_{HK} = -5.16$.  Using the 
relation between rotation period and the $S_{HK}$ index (Noyes \etal 1984) we estimate
a rotational period of 48 days for this star.  Stellar parameters are summarized 
in Table 1.

The observation dates, radial velocities and RV uncertainties for HD 88133 are listed
in Table 2.  We estimate the stellar jitter for this star to be 3.2 \ms and add this jitter  
in quadrature with the formal RV errors when fitting the data with a Keplerian and 
calculating \chisq. 
Our best fit orbital parameters are listed in Table 3 and the Keplerian fit 
is overplotted on the phased RV data in Figure 5.   The orbital period is 3.41 days,
with a velocity semi-amplitude of 35.7 \ms and eccentricity of 0.11 $\pm 0.05$ with 
RMS scatter of 5.26 \ms and \chisq of 1.26.
With the stellar mass of 1.2 \msune, we derive \msini = 0.29 \mjup and 
$a_{rel} = 0.046$ AU. 
Uncertainties were estimated by running 
100 Monte Carlo trials, fitting in the orbital parameters listed in Table 3. 
The models of Bodenheimer, Laughlin \& Lin (2003)
predict a planetary radius for HD 88133b of $0.97 R_{\rm JUP}$ if the
planet has a core, and $1.12 R_{\rm JUP}$ if it does not. 

The best fit eccentricity of 0.11 is unusual since 
giant planets with orbital periods shorter than 5 days are expected to 
be tidally circularized.  Fixing the eccentricity to zero yields 
a slightly increased \chisq of 1.48 and an RMS to the fit of 6.3 \mse. 
Again, we have added jitter in quadrature to the formal RV 
errors.  This constitutes an acceptable fit, however this star warrants additional 
RV follow-up to determine the eccentricity more accurately and to check for 
any additional planetary companions. 

In addition to the radial velocity observations, we have collected 161 
photometric measurements between 2004 March and June with the T12 0.8~m
automatic photometric telescope at Fairborn Observatory.  The transit 
probability is given by the ratio of the stellar radius to the semi-major 
axis of the orbit: $R_*/a$ = 19.5\%.  The telescope,
photometer, observing procedures, and data reduction techniques are
described briefly in Butler \etal (2004) and the references therein.
Our photometric comparison star was HD~88270 ($V$ = 6.64, $B-V$ = 0.36, F2~V),
which has been shown to be constant to 0.002 mag or better from 
intercomparison with additional comparison stars.

The 161 combined $(b+y)/2$ differential magnitudes of HD~88133 were phased
with the planetary orbital period and a time of inferior conjunction 
(computed from the orbital elements in Table~3) and plotted in Figure~6.
The standard deviation of the observations is 0.0029 mag, slightly larger
than the typical precision with this telescope, most likely because HD~88133
is somewhat fainter than most program stars.  Period analysis does not
reveal evidence of any periodicity between 1 and 60 days.  A least-squares
sine fit of the observations phased to the radial velocity period gives
a semi-amplitude of 0.00054 $\pm$ 0.00030 mag, so starspots are unlikely
to be the cause of the radial velocity variations.  The solid curve in
Figure~6 approximates the predicted transit light curve (assuming central
transits) computed from the orbital elements, the stellar radius, and
an adopted planetary radius equal to Jupiter's.  The horizontal bar
below the predicted transit window represents the approximate uncertainty
in the time of mid transit, based on Monte Carlo simulations and the
uncertainties in the orbital elements.  The predicted transit depth is only 
0.0031 mag, due to the large stellar radius.  However, the mean of the 8 
observations within the transit window agrees with the mean of the 153 
observations outside the window to within 0.0005 mag, just as expected 
from the precision of the observations.  Thus, central (non-grazing) 
transits are ruled out by our photometry.

\section{Discussion}

The N2K consortium was established to survey ``the next 2000'' closest and 
brightest high metallicity stars using a distributed observing program 
and a quick-look strategy to identify stars with RMS velocity scatter consistent 
with the presence of a hot Jupiter companion.  Over the next two years, this program 
should identify 60 new gas giant planets in short period orbits, quadrupling the number of
hot Jupiters that have been discovered and providing 
6 new transiting planets around stars brighter than $V=10.5$. 

Observations of the first 211 stars at Keck have identified 
35 stars with velocity variations consistent with the presence of an extrasolar 
planet.  Follow-up RV observations on one of these 
stars (HD 88133) has confirmed the presence of a planet in 
a 3.41 day orbit with \msini = 0.29 \mjupe.  
Our photometric observations have established that HD~88133 is highly
constant on the radial velocity period and that it does not undergo
transits at the predicted times.

Spectroscopic analysis of the program stars  
requires about one CPU hour to run, so it is possible to analyze 
$\sim 300$ stars in one 24 hour period on our computer cluster at San Francisco 
State University.  This pipeline analysis makes it possible to 
check every star the morning after the first observation is obtained.  Unsuitable 
stars (SB2s, rapid rotators) are immediately dropped and the observing list 
is backfilled with new stars.

In addition to the 250 stars observed in the 2004A semester at Keck, 
we have completed 3 or more observations of 130 stars at Subaru.  All of these stars have 
spectroscopic and Doppler analysis completed and we are obtaining follow-up observations on those 
stars with RMS velocity scatter greater than $3 \sigma$ and less than 100 \mse.  At 
Magellan, 90 stars have multiple observations and we are beginning the Doppler 
analysis of those stars.  So, the first 500 of 2000 metal-rich stars
are now beginning to be processed by the N2K consortium.

\acknowledgements

We gratefully 
acknowledge the dedication and support of the Keck Observatory staff. 
We thank the NOAO and NASA Telescope assignment committees for generous allocations
of telescope time. We thank the Michaelson Science Center for travel support 
and support through the KDPA program. GWH acknowledges support from NASA grant NCC5-511 and
NSF grant HRD-9706268. We acknowledge support by NASA grant NAG5-75005 (to GWM) and by NSF grant
AST-9988358 and NASA grant NAG5-4445 (to SSV) and NASA grant NNG04GKi9G (to GL) 
and are grateful to Sun Microsystems for ongoing support.  
This research has made use of the 
Simbad database, operated at CDS, Strasbourg, France. 
The authors extend thanks to those of native Hawaiian ancestry on whose sacred mountain 
of Mauna Kea we are privileged to be guests.  Without their generous hospitality, the Keck 
observations presented herein would not have been possible.

\clearpage

\begin{figure}
\plotone{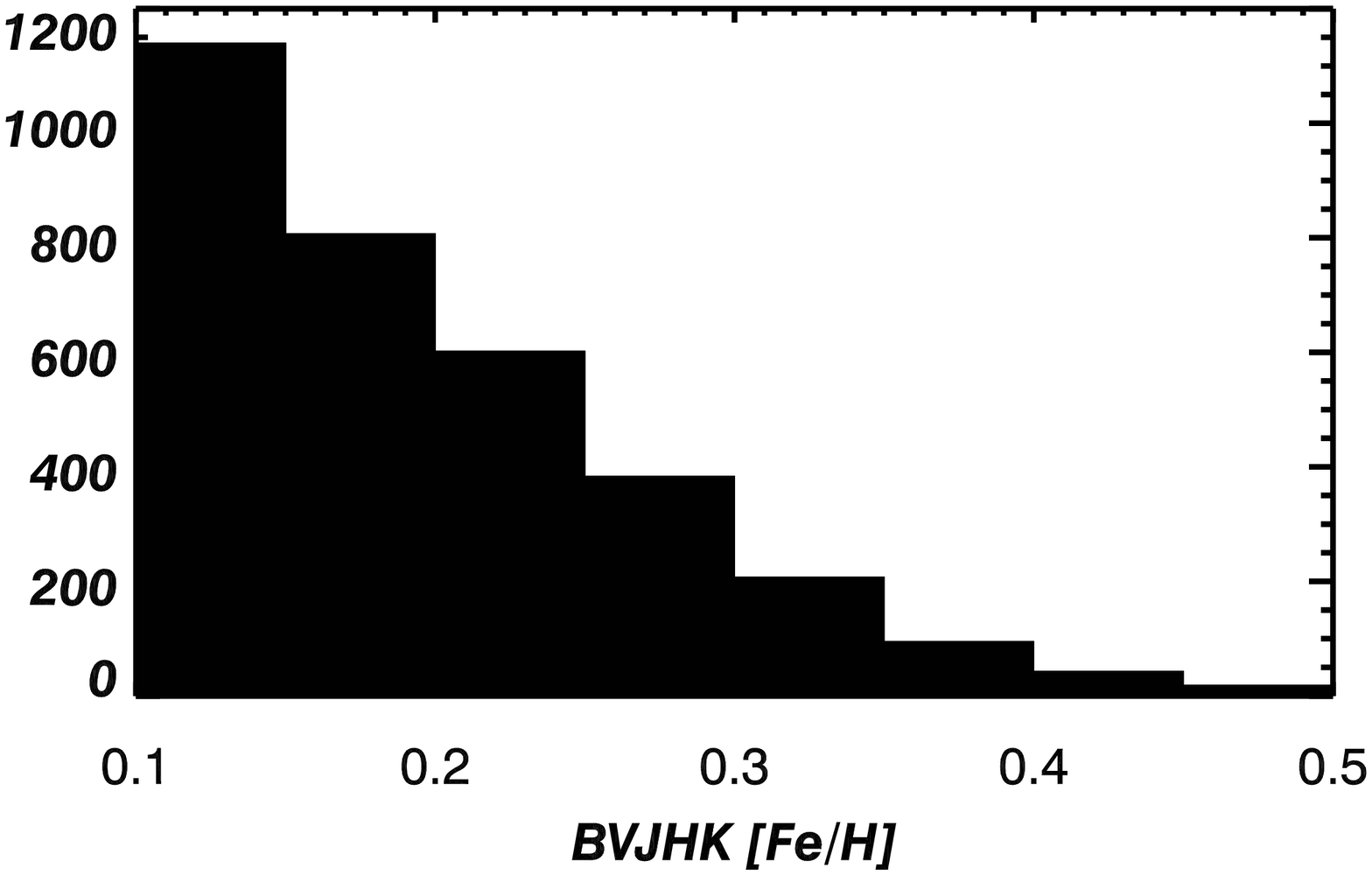}
\caption{The metal-rich tail of the [Fe/H] distribution for our N2K reservoir 
sample of more than 14,000 stars.  Broadband photometry ($B_T,V_T, JHK$) was used to estimate the 
metallicity for all main sequence and subgiant stars closer than 110 pc, brighter than $V = 10.5$, 
with color $0.4 < B-V < 1.2$ and not on current Doppler programs.  The 2000 stars shown here  
have [Fe/H] $> 0.1$ dex and are the primary targets for the N2K program}
\label{fig1}
\end{figure}
\clearpage

\begin{figure}
\plotone{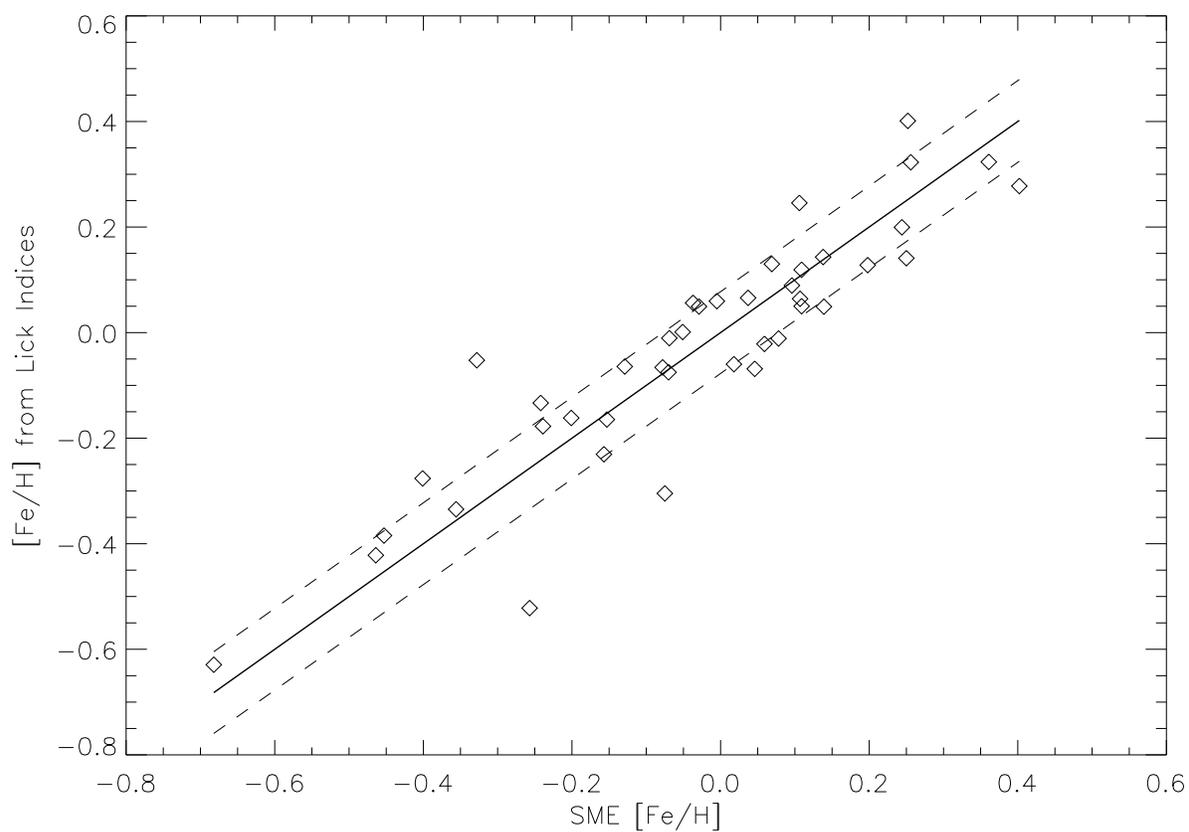}
\figcaption{Low resolution indices based on equivalent widths of a set of 
spectral lines have been calibrated to metallicities derived from spectral 
synthesis modeling of high resolution spectra, and show agreement to better 
than 0.1 dex}
\label{fig2}
\end{figure}
\clearpage

\begin{figure}
\plotone{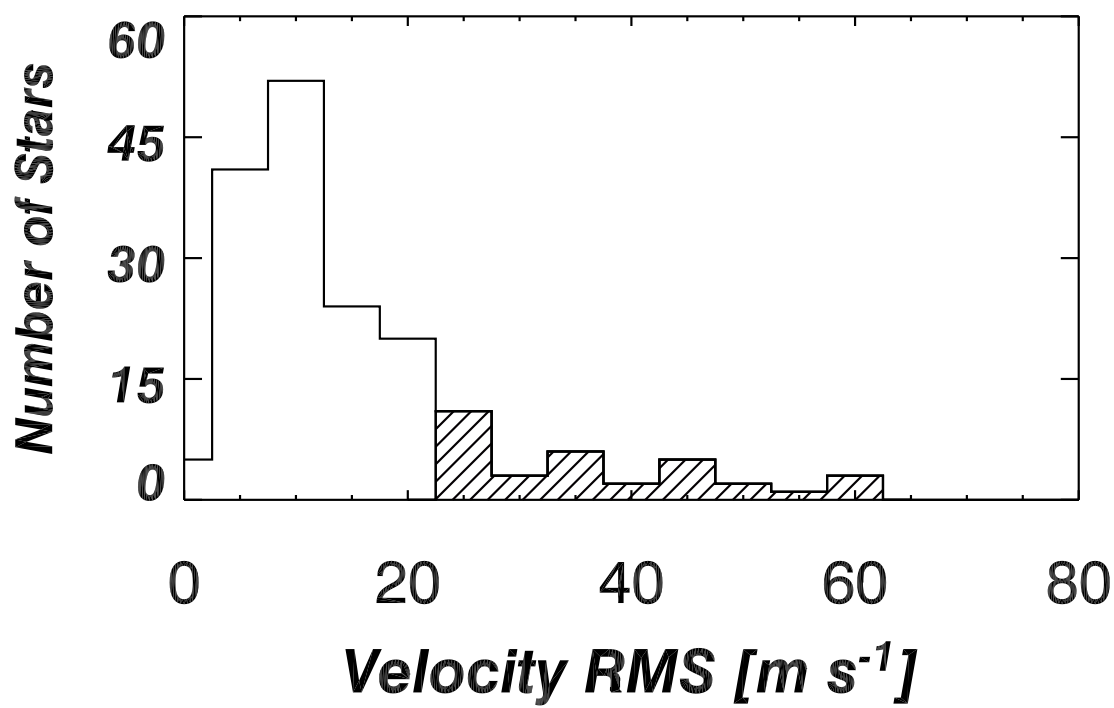}
\figcaption{Velocity scatter less than 80 \ms is observed for 175 of 211 stars
observed at Keck after 3 or more RV observations.  RV scatter $< 3 \sigma$ is seen 
for 148 of these stars. 
However 35 stars, or 16\% of the observed sample (indicated in the cross-hatched bins) 
show $> 3 \sigma$ RV scatter, consistent with the presence of a hot Jupiter. The planet 
announced in this paper had an initial velocity RMS of 24 \mse.  }
\label{fig3}
\end{figure}
\clearpage

\begin{figure}
\epsscale{0.7}
\plotone{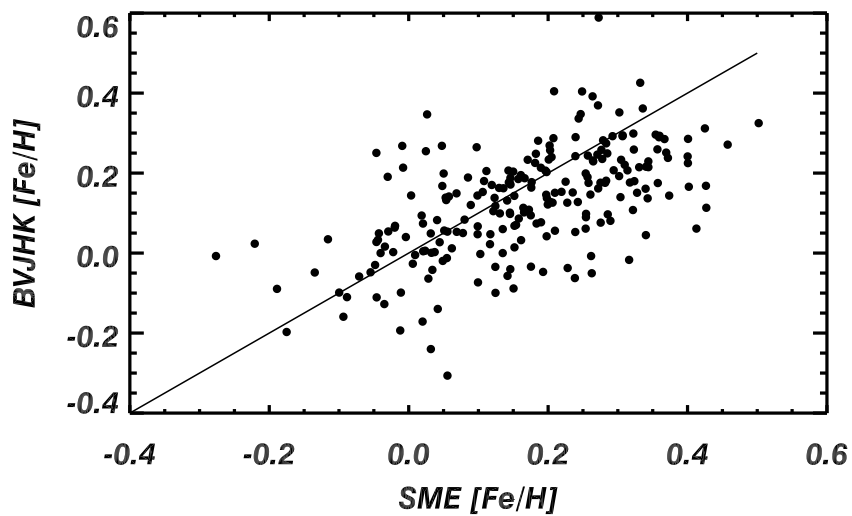}
\figcaption{A comparison of spectroscopically-derived [Fe/H] with 
the metallicity estimate from our BVJHK calibration (Ammons \etal 2004) for 
211 stars observed at Keck. The solid line shows a 1 to 1 correlation.}
\label{fig4}
\end{figure}
\clearpage

\begin{figure}
\plotone{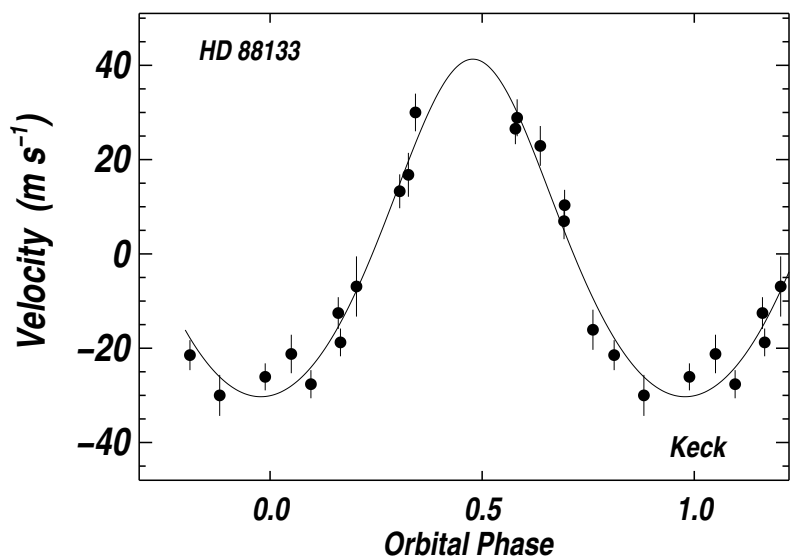}
\figcaption{Phased radial velocities for HD 88133.  
With an orbital period of 3.41 d, velocity amplitude
of 35.7 \ms and stellar mass of 1.2 \msun we derive a planet mass, 
\msini = 0.29 \mjup
and orbital radius of 0.046 AU. }
\label{fig5}
\end{figure}
\clearpage

\begin{figure}
\plotone{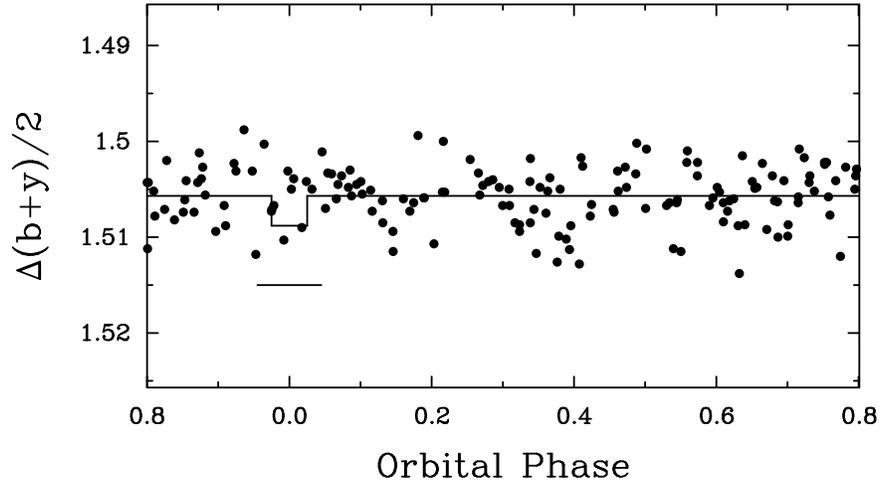}
\figcaption{ Str\"omgren $(b+y)/2$ photometric observations of HD~88133 
acquired with the T12 0.8~m APT at Fairborn Observatory and phased to the 
radial velocity period.  There is no evidence in the observations 
for any periodicity between 1 and 100 days.  The star is constant on the 
radial velocity period to a limit of 0.0005 mag, supporting the planetary 
interpretation of the radial velocity variations.  Although predicted transit 
depths are only 0.003 mag due to the large stellar radius, such transits 
are nonetheless ruled out by the photometry.}
\label{fig6}
\end{figure}
\clearpage

\begin{deluxetable}{ll}
\tablenum{1}
\tablecaption{Stellar Parameters for HD88133}
\tablewidth{0pt}
\tablehead{\colhead{Parameter}  & \colhead{} \\
} 
\startdata
V                  & 8.01          \\
$M_V$              & 3.65          \\
B-V                & 0.81          \\
Spectral Type      & G5 IV         \\
Distance (pc)      & 74.46         \\
${\rm [Fe/H]}$     & 0.34 (0.04)    \\
$T_{eff}$ (K)      & 5494 (23)      \\
\vsini \ks         & 2.2 (0.3)     \\
\logg              & 4.23 (0.05)   \\
$M_{STAR}$ (\msun) & 1.2 (0.2)    \\
$R_{STAR}$ (\rsun) & 1.93 (0.06)   \\
$S_{HK}$           & 0.138        \\
$\log R'_{HK}$     & -5.16        \\
$P_{ROT}$ (d)      & 48.0         \\
\enddata                         
\end{deluxetable}                           
\clearpage

\begin{deluxetable}{lrc}
\tablenum{2}
\tablecaption{Radial Velocities for HD88133}
\tablewidth{0pt}
\tablehead{ 
\colhead{JD}  &   \colhead{RV}    & \colhead{Uncertainties}  \\
\colhead{-2440000}  &  \colhead{(\ms)}  &   \colhead{(\ms)}  \\
}
\startdata
 13014.948 &   -14.3 &     4.09  \\
 13015.947 &    36.9 &     3.99  \\
 13016.953 &    29.8 &     4.22  \\
 13044.088 &    35.8 &     3.92  \\
 13044.869 &   -14.6 &     3.18  \\
 13045.843 &   -20.7 &     2.98  \\
 13046.081 &   -11.9 &     2.94  \\
 13069.016 &   -23.1 &     4.35  \\
 13071.788 &    13.8 &     3.82  \\
 13072.021 &    -9.2 &     4.26  \\
 13073.950 &    23.7 &     4.66  \\
 13076.948 &     0.0 &     6.38  \\
 13153.760 &    17.3 &     3.25  \\
 13179.754 &    20.2 &     3.59  \\
 13195.748 &   -19.2 &     2.86  \\
 13197.762 &    33.5 &     3.27  \\
 13199.751 &    -5.6 &     3.36  \\
\enddata
\end{deluxetable}
\clearpage

\begin{deluxetable}{ll}
\tablenum{3}
\tablecaption{Orbital Parameters for HD88133b}
\tablewidth{0pt}
\tablehead{\colhead{Parameter}  & \colhead{} \\
} 
\startdata
P (d)                    &  3.415 (0.001)  \\
${\rm T}_{\rm p}$ (JD)   &  2453016.4 (1.2) \\
$\omega$ (deg)           &  10.2 (162.9)  \\
ecc                      &  0.11 (0.05)  \\
K$_1$ (\ms)              &  35.7 (2.2)   \\
a (AU)                   &  0.046     \\
a$_1 \sin i$ (AU)        &  1.11e-05    \\
f$_1$(m) (M$_\odot$)     &  1.58e-11  \\
$M\sin i$ (M$_{Jup}$)    &  0.29       \\
${\rm Nobs}$             &  17         \\
RMS (\ms)                &  5.3       \\
Reduced \chisq           &  1.27       \\
\enddata                        
\end{deluxetable}                          
\clearpage

\end{document}